\documentstyle[12pt,aaspp4]{article}

\begin{document}

\title{An Implicit Lagrangian Code for Spherically Symmetric \\ 
General Relativistic Hydrodynamics \\
with an Approximate Riemann Solver}
\author{Shoichi Yamada\altaffilmark{1}}
\affil{Max-Planck-Institut f\"{u}r Astrophysik}
\authoraddr{Karl - Schwarzschild - Str. 1, D-85740 Garching, Germany}
\altaffiltext{1}{Department of Physics, Graduate School of Science, 
The University of Tokyo}
\authoremail{shoichi@MPA-Garching.MPG.DE}

\begin{abstract}
An implicit Lagrangian hydrodynamics code for general relativistic spherical 
collapse is presented. This scheme is based on an approximate
linearized Riemann solver (Roe type scheme) and needs no artificial
viscosity. This code is aimed especially at the calculation of the late 
phase of collapse-driven supernovae and the nascent neutron star,
where there is a remarkable contrast between the dynamical time scale
of the proto-neutron star and the diffusion time scale of neutrinos, 
without such severe limitation of the Courant condition at the center of the
neutron star. Several standard test calculations have been done
and their results show (1) this code captures the shock wave
accurately, though some erroneous jumps of specific internal energy
are found at the contact discontinuity in the shock tube problems. 
(2) The scheme shows no instability  even if we choose the Courant
number larger than 1. (3) However, 
the Courant number should be kept below $\sim 0.2$ at the shock
position so that the shock can be resolved with a few
meshes. (4) The scheme reproduces the well known analytic solutions to 
the point blast explosion, the gravitational collapse of the
uniform gas with $\gamma = 4/3$ and the general relativistic collapse
of uniform dust. Two other adiabatic simulations have also been done in
order to test the performance of the code in the context of the
collapse-driven supernovae. It is found that the time step can be
extended far beyond the Courant limitation at the center of the
neutron star. The details of the scheme and the results of these test 
calculations are discussed.  
\end{abstract}


\section{INTRODUCTION}

Since the pioneering work by Colgate and White (1966), detailed 
studies of the dynamics of collapse-driven supernovae
have been done mainly with numerical simulations (\cite{bet90}, 
\cite{mul91}, and references therein). 
One of the difficulties in so doing is the remarkable contrast between 
the dynamical time scale of the nascent neutron star ($\lesssim 1$msec)
and the diffusion time scale of neutrinos ($\sim 1$sec). If we want to 
simulate the whole scenario of the collapse-driven supernovae, these
time scales should be treated simultaneously. Since the typical time
scale of weak interactions ($\sim 10^{-7}$sec) is much smaller than 
the dynamical time
scale, the neutrino transfer has been treated in  implicit
ways in general. The hydrodynamics, however, has been calculated 
chiefly by explicit schemes, so that the time steps are restricted by
the Courant condition and a large number of integration steps are
required to simulate the late stage of collapse-driven supernovae. 
Two approaches have been conceived to overcome such severely limited
time steps. The first one is to implement an algorithm of
individual time steps in the explicit difference scheme (\cite{bww91}), 
in which 
the time steps at different positions are allowed to be different 
from one another and are determined from the local
Courant condition. The advantage of this approach is that it is easily 
applied to multi-dimensional simulations. On the other hand, the implicit
difference scheme is another possibility, since the time step is not
restricted by the Courant condition, though the required accuracy of
calculation will limit the time step. The disadvantage of
this approach is that the number of operations per each time 
step becomes much larger than in the corresponding explicit scheme,
particularly in multi-dimensional simulations. Hence, it is of no use
unless we can take time steps large enough to make up for its larger
operation number. However, note that the nascent
neutron star, whose dynamical time scale limits the time step in the
explicit schemes, evolves in a nearly hydrostatic manner so that it might 
be possible to take large enough time steps in this case. Note also
that an advantage of implicit hydrodynamic schemes in the
simulation of collapse-drive supernovae is that the 
neutrino transfer schemes are generally coded in implicit fashions so
that the hydrodynamic parts are easily incorporated into the neutrino 
parts. 
\par
The purpose of this paper is to provide such an implicit scheme 
from the latter stand point. Such attempts have already been made by
Schinder et al. (1988) and Swesty (1995). Their schemes use  
almost the same equation set as that of May \& White (1967), which has been 
frequently used in explicit general relativistic calculations
(note, however, that Schinder et al. extended it to polar-slicing 
coordinates).
Their codes also need  artificial viscosities. On the other hand,
for the present scheme we tried to apply a so-called Godunov-type
differencing method (\cite{hir90} and references therein), which has 
been extensively utilized for the
multi-dimensional hydrodynamic simulations these days. Since we have 
evaluated the pressure and velocity at a cell interface using the
solution of the linearized advection equations, this code is an 
extension of Roe's scheme. Thanks to the  numerical diffusion induced 
implicitly by finite differencing, this code needs no artificial viscosity
like most of the recent elaborate Euler schemes. In addition, it is
quite robust. In fact, as shown later, the calculation does not
collapse even if we take time steps much larger than the Courant 
limit, (though the results become less accurate). 
\par
This paper is not aiming at the general application of Roe's scheme to 
general relativistic hydrodynamics (\cite{eld95}, \cite{rom95}). 
The coordinates and the metric we use here are fully reduced ones under
the assumption of spherical
symmetry so that the degree of time slicing does not remain at all. The
equation set is also so chosen that they look similar to the nonrelativistic 
counter parts in these coordinates. Instead, the final goal of this
project is to 
study the neutrino physics in collapse-driven supernovae by combining
this code with sophisticated neutrino transfer codes. Recently, many
researchers in this field have considered that the key factors of 
successful explosions are neutrino and multi-dimensional hydrodynamical 
effects such as convection (\cite{her94}, \cite{bur95},
\cite{jan95}, \cite{shi94}). It is true that the spherical simulations cannot
treat such multi-dimensional effects properly, but we think that they can
play a complementary role to multi-dimensional simulations since 
multi-dimensional simulations use a more approximate treatment of 
neutrino transport than in spherical calculations. At
present we are trying to incorporate this hydrodynamical code into the 
multi-group flux-limited diffusion code (\cite{suz90}). 
The coding of the Boltzmann solver has also been undertaken. 
This paper is the first step of the project.  
\par
The organization of this paper is as follows. The details of the
formulations and the schemes are described in the next section. The
results of the test calculations are shown in the section
3. Concluding remarks are provided in the final section.

\subsection{Basic Equations}

Spherical symmetry of the system is assumed and the following form of
the metric is used:
\begin{equation}
ds^{2} = e^{2 \phi(t, m)}c^{2} dt^{2} - e^{2 \lambda(t, m)} 
\left(\frac{G}{c^{2}}\right)^{2} dm^{2} - r^{2}(t, m) 
(d\theta^{2} + \sin^{2}\theta d\phi^{2}) \quad .
\end{equation}
In the above formula $c$ and $G$ are the velocity of light and the
gravitational constant, respectively, which are taken to be unity in
the following equations. $t$ is the coordinate time and $m$ is the baryon mass
coordinate which is related to the circumference radius $r$ through the
conservation law of baryon mass as described below. This form of the
metric is so simple that it has been frequently adopted in the
spherically symmetric simulations thus far. However, this way of 
time slicing is not suitable for the study of black hole formation
(see \cite{sch88}). Since our purpose is to 
investigate the supernova dynamics leading to  proto-neutron star 
formation, that is not a serious problem. The basic equations consist
of the Einstein equations obtained from the above metric: 
\begin{equation}
G^{\mu \nu} = 8 \pi T^{\mu \nu}
\end{equation}
and the Euler 
equations: 
\begin{eqnarray}
\nabla_{\nu} T^{\mu \nu} & = & 0\\
T^{\mu \nu} & \equiv & \{\rho_{b}(1 + \varepsilon) + p\} u^{\mu} u^{\nu}
- p g^{\mu \nu}
\end{eqnarray}
with the baryon number conservation equation: 
\begin{equation}
\label{eqn:bary}
\nabla _{\mu} (\rho_{b} u^{\mu}) = 0
\end{equation}
and the evolution equation of electron fraction (see below). Here
$G^{\mu \nu}$ and $T^{\mu \nu}$ are the Einstein tensor and the
energy-momentum tensor, respectively. $\rho_{b}$ is the baryon mass
density and $u^{\mu}$ is the four velocity of the matter. $\varepsilon$ 
and $p$ are the specific internal energy density and the matter pressure,
respectively. $g^{\mu \nu}$ is the inverse of the metric tensor $g_{\mu 
\nu}$. In addition to these equations, the neutrino transfer equations 
should be implemented finally, but they are dropped in this paper. For
this reason, the source term which evolves the electron fraction is
not included in the present calculations. First I
will write down the hydrodynamic equations with those representing
the baryon number conservation  and the evolution of electron fraction. 
They are formulated as follows:
\begin{eqnarray}
\label{eqn:dens}
e^{-\phi} \frac{\partial \tau}{\partial t} & = & 
\frac{1}{\Gamma} \frac{\partial}{\partial m} ( 4 \pi r^{2} U) 
- \frac{4 \pi r^{2} \tau F_{\nu}}{\Gamma r}\\
e^{-\phi} \frac{\partial U}{\partial t} & = &
-\frac{\Gamma}{h} 4\pi r^{2} \left( \frac{\partial p}{\partial m} + 
\frac{\tau q}{4 \pi r^{2}} \right) - \frac{\stackrel{\sim}{m}}{r^{2}} 
- 4 \pi r (p + p_{\nu})\\
\label{eqn:simene}
e^{-\phi} \frac{\partial \varepsilon}{\partial t} & = &
- \frac{1}{\Gamma}\left\{ p \frac{\partial}{\partial m} (4 \pi r^{2}
U) - p \frac{4 \pi r^{2} \tau F_{\nu}}{r} \right\} - \tau Q  \\
& = & - \frac{1}{\Gamma} \frac{\partial}{\partial m} (4 \pi r^{2} p U) 
- \frac{h}{\Gamma^{2}} e^{-\phi} \frac{\partial}{\partial t}
(\frac{1}{2} U^{2}) + \frac{h}{\Gamma^{2}} \stackrel{\sim}{m}
e^{-\phi} \frac{\partial}{\partial t} \left(\frac{1}{r}\right)
\nonumber \\
\label{eqn:energy}
&& - \frac{h U}{\Gamma^{2}} 2 \pi e^{-\phi} \frac{\partial
r^{2}}{\partial t} (p + p_{\nu}) - \frac{1}{\Gamma}
\tau U q + \frac{p}{\Gamma} 4 \pi r \tau F_{\nu} - \tau Q\\
\label{eqn:Ye}
e^{- \phi} \frac{\partial Y_{e}}{\partial t} & = & 
- m_{u} \tau \int \left( \frac{\delta f_{\nu}}{\delta \lambda}
\right)_{\mbox{\small coll.}} \frac{d^{3} p}{p^{0}}\\
\label{eqn:rcub}
\frac{\partial (\frac{4}{3} \pi r^{3})}{\partial m} & = &
~\ \Gamma \tau \quad .
\end{eqnarray}
In the above equations, $\tau \equiv
\displaystyle{\frac{1}{\rho_{b}}}$ is the inverse of the baryon mass 
density, and
$U \equiv e^{-\phi} \displaystyle{\frac{\partial r}{\partial t}}
\equiv D_{t} r$ is the radial fluid velocity. $\Gamma$ is the general
relativistic gamma factor which is defined below (equation (\ref{eqn:gamma})) 
and is solved simultaneously. $h$ is the specific enthalpy which is 
also defined below (equation (\ref{eqn:enth})) and is treated as the dependent 
variable to be solved. $\stackrel{\sim}{m}\!(r)$ is the gravitational
mass inside the radius of $r$ and should be distinguished from the
baryon mass coordinate $m$. The second representation of the energy 
conservation (\ref{eqn:energy}) looks peculiar. This form of the
equation declares explicitly that we get the internal energy density
by solving the equation of the total energy conservation of the fluid 
(in the Newtonian limit) and then extracting  
the kinetic energy and the gravitational energy from it rather than
using the simpler form (\ref{eqn:simene}) which is just the first law
of thermodynamics. Both forms of the energy 
equation are tried in the shock tube calculations and it is found that 
the latter form (\ref{eqn:energy}) better reproduces the 
Rankine-Hugoniot relation. This is a well known fact in the Eulerian
numerical simulations. Note that the artificial viscosity 
is not necessary in both cases for the finite difference scheme
described below. $F_{\nu}$, $p_{\nu}$, $q$, $Q$ and $f_{\nu}$ are
quantities related with neutrinos. They are defined as follows:
\begin{eqnarray}
T_{\nu} ^{\alpha \beta} & \equiv & 
\rho_{b} E_{\nu} u^{\alpha} u^{\beta}+ F_{\nu}^{\alpha} u^{\beta}
+ u^{\alpha} F_{\nu} ^{\beta} + p_{\nu}^{\alpha \beta}\\
Q^{\alpha} & \equiv &
T_{\nu}^{\alpha \beta}{} _{; \alpha} \equiv (e^{-\phi}Q, e^{-\lambda}q,
0, 0) \\ 
\label{eqn:Qa}
& = & \int \left( \frac{\delta f_{\nu}}{\delta \lambda } \right)
_{\mbox{\small coll.}} p^{\alpha} \frac{d^{3}p}{p^{0}} \quad .  
\end{eqnarray}
$T_{\nu}^{\alpha \beta}$ is the energy-momentum tensor of neutrinos
and $\rho_{b}$ is the baryon mass density. $E_{\nu}$,
$F_{\nu}^{\alpha}$ and $p_{\nu}^{\alpha \beta}$ are, respectively, the 
energy density, the flux vector and the stress tensor of
neutrinos. The right hand sides of equations (\ref{eqn:Ye}) and
(\ref{eqn:Qa}) are the phase space integrals of the collision term in the 
neutrino Boltzmann equations.  
Although these terms are included in the above equations for
completeness, they are omitted in the present calculations. In
particular, the right hand side of equation (\ref{eqn:Ye}) is set to be
zero, so that $Y_{e} = \mbox{const.}$ in all of the computations below.  
Equation (\ref{eqn:rcub}) relates the radius and the baryon mass inside
it and is obtained from the combination of the baryon number conservation 
equation (\ref{eqn:bary}) and the definition of $\Gamma$ 
below (equation (\ref{eqn:gamma})). This
equation is used to determine $r(m)$ of the next step instead of $D_t
r = U$, which has been often used so far. It should be noted that
the above equations are very similar to the Newtonian counterparts. As 
a result, the difference scheme developed for the Newtonian system can 
be applied to the above general relativistic system. The rest of the
basic equations specify the metric components $\lambda(t, m)$ and
$\phi(t, m)$, the gamma factor $\Gamma(t, m)$, the gravitational
mass $\stackrel{\sim}{m}(t, m)$ and the specific enthalpy $h(t,
m)$. These equations are:
\begin{eqnarray}
\label{eqn:lmbd}
e^{\lambda} & = & ~\ \frac{1}{\Gamma}\frac{\partial \, r}{\partial m}\\
\label{eqn:gamma}
\Gamma^{2} & = & ~\ 1 + u^{2} - \frac{2 \! \stackrel{\sim}{m}}{r} \\
\label{eqn:gravpot}
h \frac{\partial \phi}{\partial m} & = & 
- \tau \frac{\partial \, p}{\partial m} - \frac{\tau^{2}q}{4 \pi r^{2}}\\
\label{eqn:gravm}
\frac{\partial \stackrel{\sim}{m}}{\partial t} & = & 
- (p + p_{\nu}) \frac{\partial}{\partial t} (\frac{4}{3} \pi r^{3})
- 4 \pi r^{2} e^{\phi} \Gamma F_{\nu}\\
\label{eqn:enth}
h & = & ~\ 1 + \varepsilon + p \, \tau  \quad .
\end{eqnarray} 
As is clear, these are not evolutionary equations except for equation
(\ref{eqn:gravm}), but are constraint equations which must be
satisfied at each time step. Since the present scheme is written in the
implicit form, these constraints are always satisfied within the
error for convergence of iteration. The equation for the
gravitational mass $\stackrel{\sim}{m}$ can be written in the
constraint form as well:
\begin{equation}
\label{eqn:const}
\frac{\partial \! \stackrel{\sim}{m}}{\partial m} = 
4 \pi r^{2} \, [\, \rho_{b} \, (1 + \varepsilon + E_{\nu}) + \frac{U
F_{\nu}}{\Gamma}]\, \frac{\partial \, r}{\partial m} \quad .
\end{equation}
However, equation (\ref{eqn:gravm}) is adopted here rather than
(\ref{eqn:const}), since it explicitly guarantees the total energy
conservation $\stackrel{\sim}{m}\!(t, R) = \mbox{const.}$, where $R$ is
the stellar core radius, as long as the energy loss due to the emission 
of neutrinos is disregarded. The constraint equation (\ref{eqn:const}) 
is used for the check of the accuracy of calculations. The terms
concerning the interactions of neutrinos are included in the above
equations also, but are dropped in the present simulations again. Equation
(\ref{eqn:enth}) is nothing but the definition of specific enthalpy. 
Usually, this expression is expanded in the preceding
equations. However, it is treated as one of the dependent variables to 
be solved simultaneously in this code, since it avoids the repeated
appearance of the derivatives of $h$ in the matrix elements. This
treatment means that the specific enthalpy does not satisfy the above
definition strictly during the iteration process. This did not cause any
serious problem in the test calculations shown below. Moreover it
increases CPU time very little. It should be noticed that the most  
time consuming part in the radiative hydrodynamics code is the neutrino transfer
subroutine. In order to integrate equation (\ref{eqn:gravpot}) we must 
specify the boundary condition for $\phi$. It is done so that the
metric in the stellar core coincides at the surface of the stellar
core with the Schwarzschild metric for a spherically symmetric vacuum:
\begin{equation}
\label{eqn:bndc}
e^{\phi_{s}} = \left( 1 - \frac{2 \! \stackrel{\sim}{m}_{s}}{r_{s}}
\right ) \left( 1 + U_{s}^{2} - \frac{2 \!
\stackrel{\sim}{m}_{s}}{r_{s}} \right)^{-\frac{1}{2}}\; ,
\end{equation}
where the subscript $''s''$ means the variables are evaluated at the
surface of the stellar core. Strictly speaking, the metric outside the 
stellar core is not the Schwarzschild metric because neutrinos
exist there. For simplicity, however, the above approximation is
adopted in the following calculations. To summarize, the basic
equations to be solved are equations (\ref{eqn:dens}) through 
(\ref{eqn:rcub}) and (\ref{eqn:lmbd}) through (\ref{eqn:enth}). The
latter form (\ref{eqn:energy}) is used for the energy conservation
equation as stated above. The boundary condition is given as equation 
(\ref{eqn:bndc}). These equations are differenced in the implicit
method described in the next section. The resulting equations are the
simultaneous equations consisting of $10 \times N$ (mesh number)
equations. The matrix to be inversed has a block-tridiagonal form.  

\subsection{The Implicit Differencing Scheme}

All the dependent variables except $r$, $\phi$ and
$\stackrel{\sim}{m}$ are defined at the mesh centers as in most 
of the recent Eulerian schemes. The values at the cell interfaces 
for pressure and velocity are calculated by using the solutions of the 
linearized Riemann problems as described below.  $r$, $\phi$ 
and $\stackrel{\sim}{m}$ are placed at
the edges of meshes in order to simplify the integrals 
(equation (\ref{eqn:gravpot})) and the
implementation of the boundary condition (equation (\ref{eqn:bndc})). 
The superscripts and the subscripts attached to the variables 
in the following difference equations are representing the time step 
and the mesh number, respectively. It is assumed that the mesh centers 
are specified by the lower case subscript like $''i''$, on the other
hand to the interfaces attached are the uppercase subscript like
$''I''$, and that all these subscripts are integers. It follows that 
$i$-th mesh center ($i = 1\sim N$) is located between the $(I-1)$-th 
and $I$-th interfaces ($I = 1\sim N$). However, there is one
exception. $\phi_{I}$'s are defined on the $(I-1)$-th interfaces. This 
is because the boundary condition of $\phi$ (equation
(\ref{eqn:bndc})) is specified on the surface of the iron core ($N$-th 
interface). This is not the case for 
$r$ and $\stackrel{\sim}{m}$, for which the inner boundary conditions: 
$r_{c} = 0$ and $\stackrel{\sim}{m}_{c} = 0 $ are assumed implicitly
and we don't have to solve them. 
\par
Paying  attention to these assumptions, we will first see the 
finite differenced equations regarding the $i$-th mesh and $I$-th 
interface (but $(I-1)$-th interface only for $\phi$):
\begin{eqnarray}
\label{eqn:voldif}
0 & = &\frac{\Delta m_{i}}{4 \pi \Delta t} e^{-\ll \phi \gg^{n+\frac{1}{2}}
_{i}}\left(\tau^{n+1}_{i} - \tau^{n}_{i}\right) 
- \frac{1}{< \Gamma > ^{n+\frac{1}{2}}_{i}} \left\{ 
\left( < r >^{n+\frac{1}{2}}_{I} \right)^{2} u^{n+1}_{I} - 
\left( < r >^{n+\frac{1}{2}}_{I-1} \right)^{2} u^{n+1}_{I-1}
\right\}
\nonumber \\
&& \\
0 & = & \frac{\Delta m_{i}}{4 \pi \Delta t} e^{-\ll \phi \gg^{n+\frac{1}{2}}
_{i}}\left(U^{n+1}_{i} - U^{n}_{i}\right) 
+ \frac{<\Gamma>^{n+\frac{1}{2}}_{i}}
{<h>^{n+\frac{1}{2}}_{i}}\left\{ 
\overline{<r>}^{n+\frac{1}{2}}_{i} \right \}
^{2} (p^{n+1}_{I} - p^{n+1}_{I-1})
\nonumber\\
& + &\frac{\Delta m_{i}}{4 \pi} \frac{\ll \stackrel{\sim}{m}
 \gg^{n+\frac{1}{2}}_{i}}
{\overline{r}^{n}_{i} \overline{r}^{n+1}_{i}} 
+ \Delta m_{i} \overline{<r>}^{n+\frac{1}{2}}_{i}
<p>^{n+\frac{1}{2}}_{i}
\end{eqnarray}
\newpage
\begin{eqnarray}
0 &=& \frac{\Delta m_{i}}{4 \pi \Delta t} e^{-\ll \phi \gg^{n+\frac{1}{2}}
_{i}}\left(\varepsilon^{n+1}_{i} - \varepsilon^{n}_{i}\right) 
\nonumber \\
&+& \frac{1}{<\Gamma>^{n+\frac{1}{2}}_{i}} 
\left\{ \left(<r>^{n+\frac{1}{2}}_{I}\right)^{2} u^{n+1}_{I}
p^{n+1}_{I} 
- \left(<r>^{n+\frac{1}{2}}_{I-1}\right)^{2} u^{n+1}_{I-1}
p^{n+1}_{I-1} \right\}
\nonumber \\
&+&  \frac{<h>^{n+\frac{1}{2}}_{i}}{\Gamma^{n}_{i}\Gamma^{n+1}_{i}}
e^{-\ll \phi \gg^{n+\frac{1}{2}}_{i}} \frac{\Delta m_{i}}{4 \pi \Delta 
t}\frac{1}{2}\left\{\left(U^{n+1}_{i}\right)^{2} 
- \left(U^{n}_{i}\right)^{2}\right\}
\nonumber \\
&-&  \frac{<h>^{n+\frac{1}{2}}_{i}}{\Gamma^{n}_{i}\Gamma^{n+1}_{i}}
e^{-\ll \phi \gg^{n+\frac{1}{2}}_{i}} \frac{\Delta m_{i}}{4 \pi \Delta 
t}\ll \stackrel{\sim}{m} \gg^{n+\frac{1}{2}}_{i}
\left( \frac{1}{\overline{r}^{n+1}_{i}} 
- \frac{1}{\overline{r}^{n+1}_{i}} \right)
\nonumber \\
&+&  \frac{<h>^{n+\frac{1}{2}}_{i}}{\Gamma^{n}_{i}\Gamma^{n+1}_{i}}
\frac{\Delta m_{i}}{\Delta t} e^{-\ll \phi \gg^{n+\frac{1}{2}}_{i}} 
\frac{1}{2} \left\{ \left( \overline{r}^{n+1}_{i} \right)^{2}
- \left(\overline{r}^{n}_{i} \right)^{2} \right \}
<p>^{n+\frac{1}{2}}_{i}\\
0 & = &\frac{\Delta m_{i}}{4 \pi \Delta t} e^{-\ll \phi \gg^{n+\frac{1}{2}}
_{i}}\left(Y_{e}^{n+1}{}_{i} - Y_{e}^{n}{}_{i}\right)\\
0 &=& \frac{4\pi}{3}\left\{\left(r^{n+1}_{I}\right)^{3}
- \left(r^{n+1}_{I-1}\right)^{3}\right\} -
\Gamma^{n+1}_{i}\tau^{n+1}_{i} \Delta m_{i}\\
0 &=& e^{\lambda^{n+1}_{i}} - \frac{1}{\Delta m_{i}\Gamma^{n+1}_{i}} 
\left(r^{n+1}_{I} - r^{n+1}_{I-1}\right)\\
0 &=& \left(\Gamma^{n+1}_{i}\right)^{2} 
- \left\{ 1 + \left(U^{n+1}_{i}\right)^{2} 
- \frac{\stackrel{\sim}{m}^{n+1}_{I} + \stackrel{\sim}{m}^{n+1}_{I-1}}
{\overline{r}^{n+1}_{i}}  \right\}\\
0 &=& h^{n+1}_{i} \left( \phi^{n+1}_{I+1} - \phi^{n+1}_{I} \right)
+ \tau^{n+1}_{i}\left( p^{n+1}_{I} - p^{n+1}_{I-1} \right) \\
0 &=& \stackrel{\sim}{m}^{n+1}_{I} - \stackrel{\sim}{m}^{n}_{I}
+ \ p^{n+1}_{I} \  \frac{4\pi}{3}\left\{ (r^{n+1}_{I})^{3} 
- ( r^{n}_{I} )^{3} \right\} \\
\label{eqn:entdif}
0 &=& h^{n+1}_{i} - \left( 1 + \varepsilon^{n+1}_{i} 
+ \tau^{n+1}_{i} p^{n+1}_{i} \right) \quad .
\end{eqnarray}
In the above equations it is supposed that the time step $''n''$ 
corresponds to  the present time, where all the values of the
dependent variables are known. The above finite differenced equations form
nonlinearly coupled simultaneous equations. The Newton-Raphson scheme
is used to solve these equations. Some averaging procedures are
abbreviated like $<\cdots>$, $\ll \cdots \gg$ and
$\overline{\cdots}$. Their definitions are as follows:
\begin{eqnarray}
<X>^{n+\frac{1}{2}}_{i} & \equiv & \frac{X^{n+1}_{i} + X^{n}_{i}}{2}\\
\ll Y \gg^{n+\frac{1}{2}}_{i} & \equiv & \frac{Y^{n+1}_{I} +
Y^{n+1}_{I+1} + Y^{n}_{I} + Y^{n}_{I+1}}{4}\\
\overline{Z}^{n}_{i+\frac{1}{2}} & \equiv & 
\left\{ \frac{\left(Z^{n}_{I}\right)^{3} +
\left(Z^{n}_{I-1}\right)^{3}} {2}  \right\}^{\frac{1}{3}} \quad ,
\end{eqnarray}
where $X$ is assumed to be defined at the cell center or at the cell
interface while $Y$ is supposed to be defined at the cell edge. $Z$ is 
actually the radius $r$. Note that we  distinguish 
two velocities $U^{n+1}_{i}$ and $u^{n+1}_{I}$ and two pressures 
$p^{n+1}_{i}$ and $p^{n+1}_{I}$, the latter of which are determined from 
the linearized Riemann problems which are
evaluated at the time step $''n+1''$, where the values of the
dependent variables should be solved. That is, we consider the Riemann 
problems at each cell interface $''I''$ at each time step $''n+1''$,
the left and right constant states of which are given by
($\tau_{i}^{n+1}$, $U_{i}^{n+1}$, $p_{i}^{n+1}$) and 
($\tau_{i+1}^{n+1}$, $U_{i+1}^{n+1}$, $p_{i+1}^{n+1}$),
respectively. We solve these problems approximately by linearizing the 
advection equations as described below. It should be noted again
that these constant states are evaluated at the time step $''n+1''$, so 
that we cannot know them a priori. Hence we must solve these Riemann
problems at each iteration step of Newton-Raphson methods. The
physical meaning of the interface values obtained in this way is
rather obscure compared with the counterparts in explicit schemes,
since the exact solution of the Riemann problem describes the
interaction of the nonlinear wave {\it after} the time of the initial 
condition (the time step $''n+1''$). The interpretation here is that
they approximate the states after the interactions of nonlinear
waves during the time steps $''n''$ to $''n+1''$. Practically, these
values are given as follows. If we pay 
attention only to the advection terms, then we get
\begin{equation}  
\label{eqn:jacob}
\frac{\partial}{\partial t}
\left\{
\begin{array}{r}
\tau \\
U \\
p
\end{array}
\right \}
=  
\left(
\begin{array}{ccc}
0 & \displaystyle{\frac{e^{\phi}}{\Gamma}} 4 \pi r^{2} & 0 \\
0 & 0 & - \displaystyle{\frac{e^{\phi} \Gamma}{h}} 4 \pi r^{2} \\
0 & - \gamma \displaystyle{\frac{p}{\tau}}
\displaystyle{\frac{e^{\phi}}{\Gamma}} 4 \pi r^{2} & 0
\end{array}
\right) \frac{\partial}{\partial m}
\left\{
\begin{array}{r}
\tau \\
U \\
p
\end{array}
\right \} \quad ,
\end{equation}
where $\gamma$ is the adiabatic index and is defined as:
\begin{equation}
\gamma \equiv \left(\frac{\partial \ln p}{\partial \ln \rho_{b}}
\right)_{s} \quad\quad (\mbox{$s$ is the entropy per baryon}) \quad .
\end{equation}
In equation (\ref{eqn:jacob}), the dependent variable is changed from
$\varepsilon$ to $p$, for two reasons. Firstly, the eigen vectors
are very simple in this representation:\\
\parbox{5cm}{
\begin{eqnarray*}
\lambda_{1} & = & 
\begin{array}{c}
~\\
- e^{\phi} 4 \pi r^{2} \rho_{b} c_{s}\\
~
\end{array}\\
\lambda_{2} & = & 
\begin{array}{c}
~\\
~\ 0 \\
~
\end{array}\\
\lambda_{3} & = &   
\begin{array}{c}
~\\
~\ e^{\phi} 4 \pi r^{2} \rho_{b} c_{s}\\
~
\end{array}
\end{eqnarray*}
\hfill }
\parbox{5cm}{
\begin{eqnarray*}
\mbox{\boldmath$ r_{1}$} & = &
\left( 
\begin{array}{l}
~\ \tau \\
~\ \Gamma c_{s} \\
- \gamma p
\end{array}
\right)\\
\mbox{\boldmath$ r_{2}$} & = &
\left( 
\begin{array}{l}
~\ 1\ ~ \\
~\ 0\ ~ \\
~\ 0\ ~
\end{array}
\right)\\
\mbox{\boldmath$ r_{3}$} & = &
\left( 
\begin{array}{l}
~\ \tau \\
- \Gamma c_{s} \\
- \gamma p
\end{array}
\right)
\end{eqnarray*} 
\hfill } 
\parbox{6.3cm}{
\begin{eqnarray}
\mbox{\boldmath$l_{1}$} & = & 
\left( 
\begin{array}{l}
~\ 0\\
\displaystyle{\frac{1}{2 \Gamma c_{s}}} \\
\displaystyle{\frac{1}{2 \gamma p}}
\end{array}
\right) \\
\mbox{\boldmath$l_{2}$} & = & 
\left( 
\begin{array}{l}
~\ 1 \ ~\\
~\ 0 \ ~\\
~\,\displaystyle{\frac{\tau}{\gamma p}}
\end{array}
\right) \\
\mbox{\boldmath$l_{3}$} & = & 
\left( 
\begin{array}{l}
~\ ~\ 0 \ ~\\
-\displaystyle{\frac{1}{2 \Gamma c_{s}}} \\
-\displaystyle{\frac{1}{2 \gamma p}}
\end{array}
\right) \ ,
\end{eqnarray}
\hfill }\\
where $\lambda$'s are the eigen values, and $\mbox{\boldmath$r$}$'s
and $\mbox{\boldmath$l$}$'s are the corresponding right and left eigen 
vectors. In this equations, $c_{s}$ is the sound velocity defined as:
\begin{equation}
c_{s} \equiv \sqrt{\gamma \displaystyle{\frac{p}{\rho_{b}h}}} \quad .
\end{equation}
As in the Newtonian hydrodynamics, there are three modes, two of which 
correspond to the left- and right-going sound waves and the rest of
which represents the contact discontinuity. In this representation,
it is clear that pressure and velocity are continuous across the
contact discontinuity. This feature is taken over to the solution of
the linearized Riemann problems when these eigenvectors are used to
solve them. It is not the case when other dependent variables are
chosen. This is the second reason to use these variables. In order to
linearize equation (\ref{eqn:jacob}) we just replace the Jacobian
matrix by some constant matrix at each cell interface. In this paper, 
arithmetically averaged 
$\rho_{bm} \equiv (\rho_{bL} + \rho_{bR}) / 2$, $U_{m} \equiv (U_{L} + 
U_{R}) / 2$ and $\varepsilon_{m} \equiv (\varepsilon_{L} +
\varepsilon_{R}) / 2$ are used to evaluate the constant matrix, where
the subscripts $L$ and $R$ mean that they are evaluated at the left
and right constant states of each Riemann problem, respectively. 
The eigen values and the corresponding left and right eigen
vectors given above are also evaluated in terms of these averaged
variables at each cell interface. The linearized advection equations
thus obtained can be solved analytically. $u_{I}$ and
$p_{I}$ evaluated in this way prove to be as follows:
\begin{eqnarray}
\label{eqn:velint}
u_{I} & = & \frac{1}{2}(u_{R} + u_{L}) - \frac{c_{s m}}{2 \gamma_{m} p_{m}} 
(p_{R} - p_{L})\\
\label{eqn:presint}
p_{I} & = & \frac{1}{2}(p_{R} + p_{L}) - \frac{\gamma_{m} p_{m}}{2
c_{s m}} (u_{R} - u_{L}) \quad ,
\end{eqnarray}
where the subscript $m$ means they should be evaluated by using the above
averaged variables. It should be noted that $p_{m}$ is different from
the arithmetic mean of $p_{L}$ and $p_{R}$. As is clear, both $u_{I}$ and
$p_{I}$ are the arithmetic mean of the left- and right-state values
with the correction terms which are proportional to $(p_{R} - p_{L})$
and $(u_{R} - u_{L})$, respectively. These correction terms serve as a 
numerical diffusion so that no artificial viscosity is necessary. The
above representations of the velocity and the pressure at the cell
interface are so simple that it is quite easy to evaluate their
derivatives with respect to the dependent variables in the course
of the Newton-Raphson scheme.  
\par 
In the present code, in order to achieve spatial second order
accuracy, the piecewise linear distributions of dependent variables
$\rho_{b}$, $U$, $\varepsilon$ and $\Gamma$ are introduced. It is well 
known that some limiting procedure is necessary to maintain the
monotonicity of the numerical solution. In this paper, we use the
simplest slope prescription: if $\mbox{\boldmath $s$}_{l}$ and 
$\mbox{\boldmath $s$}_{r}$ denote the left-
and right-hand difference quotients of the above dependent variables,
the slopes $\mbox{\boldmath $S$}$  are determined by the formula
\begin{equation}
S^{i} = \left\{
\begin{array}{ll}
s^{i}_{l} & \mbox{for $\left | s^{i}_{l} \right| \le 
\left | s^{i}_{r} \right|$ and $s^{i}_{l} \cdot s^{i}_{r} > 0$} \\
s^{i}_{r} & \mbox{for $\left | s^{i}_{l} \right| >  
\left | s^{i}_{r} \right|$ and $s^{i}_{l} \cdot s^{i}_{r} > 0$} \\
0 & \mbox{otherwise \quad \quad,}
\end{array}
\right .
\end{equation} 
where the superscript $i$ distinguishes the dependent variables $\rho_{b}$, 
$U$, $\varepsilon$ and $\Gamma$. This prescription will need some
improvement in the future.

\subsection{Numerical Implementation of the Implicit Scheme}

The above finite differenced equations (\ref{eqn:voldif}) through 
(\ref{eqn:entdif}) with the definitions of interface values 
(\ref{eqn:velint}) and (\ref{eqn:presint}) form the system of the nonlinearly
coupled simultaneous equations with respect to the dependent variables 
$\tau$, $U$, $\varepsilon$, $Y_{e}$, $r$, $\lambda$, $\Gamma$, $\phi$, 
$\stackrel{\sim}{m}$ and $h$. In order to solve these equations, the
standard Newton-Raphson iteration scheme is utilized for the
linearized equations. That is, if we define the dependent variables
vector as 
\begin{equation}
\mbox{\boldmath$X$} = 
\left(
\cdots , \, 
\tau_{i},\,  U_{i},\,  \varepsilon_{i},\,  Y_{e}{}_{i},\,  r_{I},\, 
\lambda_{i},\,  \Gamma_{i},\, 
\phi_{I},\,  \stackrel{\sim}{m}_{I},\,  h_{i},\,
\cdots
\right)^{t} \quad,
\end{equation}
where the subscripts $i$ and $I$ mean the $i-$th mesh and interface
values, and denote all the equations in the vector representations as 
\begin{equation}
\mbox{\boldmath $F$}(\mbox{\boldmath $X$}) = 0 \quad ,
\end{equation}
then the linearized equations become
\begin{equation}
\label{eqn:lnr}
\mbox{\boldmath $F$}(\mbox{\boldmath $\stackrel{\sim}{X}$}) 
 + \frac{\partial \mbox{\boldmath $F$}
(\mbox{\boldmath $\stackrel{\sim}{X}$})}
{\partial \mbox{\boldmath $\stackrel{\sim}{X}$}}
\cdot \delta \mbox{\boldmath $X$} = 0 \quad,
\end{equation}
where $\mbox{\boldmath $\stackrel{\sim}{X}$}$ is some guess for
the correct $\mbox{\boldmath $X$}$. This linearized equation
(\ref{eqn:lnr}) is solved repeatedly while improving the guess by
\begin{equation}
\mbox{\boldmath $\stackrel{\sim}{X}$}_{\mbox{new}} = 
\mbox{\boldmath $\stackrel{\sim}{X}$} + \delta \mbox{\boldmath $X$}
\end{equation}
until the error $|\delta \mbox{\boldmath $X$}|$ becomes less than some
small value which is set to $10^{-5}$ in the following
calculations. As is clear, the Jacobian matrix $\displaystyle
{\frac{\partial \mbox{\boldmath $F$}
(\mbox{\boldmath $\stackrel{\sim}{X}$})}
{\partial \mbox{\boldmath $\stackrel{\sim}{X}$}}}$
turns out to be a block-tridiagonal matrix whose block size is  
$10 \times 10$. At each step of the above iteration, the linearized 
equations (\ref{eqn:lnr}) are solved by the standard Gauss elimination 
scheme, though other elaborate schemes can be applied 
to this type of matrix. Note, however, that the
matrix for the hydrodynamic sector is very small compared with that
for neutrinos. The implementation of such an efficient solver will be 
done at the next step, where the neutrino transfer code will be combined. 
\par
As stated above, the formulae for the velocity  and the pressure at
the cell interface (equation (\ref{eqn:velint}) are so simple that 
we can calculate their derivatives with regard to the dependent
variables analytically. In so doing, however, the contributions from 
the sloping prescription are dropped, which does not cause serious 
problems for convergence of the Newton-Raphson iteration process. 
The thermodynamic derivatives are obtained by differencing the EOS
table numerically, except for the adiabatic index $\gamma$, which 
is stored in the table. 

\subsection{Time-Step Control}

The time step is determined so that the change of any dependent
variables should not exceed an appropriate upper limit, which is chosen 
to be 2\% in most of the following calculations, at each time step. 
In order to save the computation time, however, we correct the next 
time step and don't repeat the same time step even though the above 
criterion is not satisfied at some time step. In fact, if the change 
of some variable exceeds the above criterion, the subsequent time step 
is decreased by a factor of 0.95. On the other hand, if none of the 
variations of the dependent variables exceeds the upper limit, then we 
multiply the next time step by 1.05. The time variation of the Courant 
number which is defined as:
\begin{equation}
\mbox{Courant number} \equiv \frac{\Delta t}{\left ( 
\displaystyle{\frac{\Delta x}{c_{s} 
+ \left | U \right|}} \right)}
\end{equation}
is monitored for all the calculations and is shown for some test
calculations below.

\section{TEST CALCULATIONS}

In the following, some results for the representative test problems are 
shown in order to clarify the performance of the present
hydrodynamical code. Among them there are shock tube problems both for 
non-relativistic and special relativistic cases, Sedov's point
explosion problem in the uniform matter, Oppenheimer-Snyder's dust 
collapse problem and the non-relativistic self-similar collapse of a
$\gamma = 4/3$ polytrope. Moreover, the results of the hydrostatic
calculations and the simple adiabatic collapse simulations are
demonstrated in order to judge whether the present code can be applied 
to the simulation for the late time stage of the collapse-driven
supernovae. In order to compare the results, the parameter sets adopted
here for the first five problems are almost the same as those studied 
by Swesty. In the course of these simulations no special tuning  
to the specific problem is done except for the boundary conditions, 
the mesh system and the equation of state.
 
\subsection{The Non-Relativistic Shock Tube Problem}

The most standard parameter set for this problem is that proposed by
Sod (1978):
\begin{equation}
\left \{
\begin{array}{ccl}
\rho_{L} & = & 1\\
p_{L} & = & 1 \\
U_{L} & = & 0
\end{array}
\right .
\quad \quad
\left \{
\begin{array}{ccl}
\rho_{R} & = & 0.125\\
p_{R} & = & 0.1 \\
U_{R} & = & 0
\end{array}
\right .
\quad \quad ,
\end{equation}
where the subscripts $L$ and $R$ mean the left- and right-state values 
and CGS units are assumed. The equation of state for this problem
is $p = (\gamma - 1) \rho_{b} \varepsilon$ with $\gamma = 4/3$ and
$5/3$. 
\par
Since the present code assumes spherically symmetric geometry, we
formulate the problem, which is originally for plane symmetric
geometry, over a thin shell whose curvature is safely
ignored. Following Swesty, we take $R = 10^{4}$cm and $\delta R = 2$cm 
for the radius and the thickness of the thin shell, respectively, and
 we define $x = R - r$ and specify the region  
$-1 < x \leq 0$ as the left state. 100 uniform meshes are 
used. Although the calculations have been done both for $\gamma = 4/3
$ and $5/3$, only the results for $\gamma = 5/3$ are described 
since the features of the numerical solutions are nearly the same 
both qualitatively and quantitatively. 
\par
In figure~1 the representative distributions of  
$\rho_{b}$, $U$, $p$ and $\varepsilon$ for $\gamma = 5/3$ are shown
with the corresponding analytic solutions. The time is 0.675sec 
(400 time steps).  
As can be seen, the shock is resolved by 3 meshes in the present
case. Moreover, no oscillations occur behind the shock and the
rarefaction wave. The Rankine-Hugoniot condition is strictly satisfied 
without any artificial viscosity. However, we can see the small jump 
of the specific
internal energy density ahead of the contact surface. This is not
an oscillation but an erroneous contact surface. This can be
understood from the smoothness of the velocity and the pressure at the 
contact surface. The approximate solution of the Riemann problem is 
least accurate at the first time step, when the largest
discontinuities exist. The observed incorrect jump of the specific
internal energy density at the contact surface is the result of the 
inaccurate approximate solution at the first time step. Since this
scheme is Lagrangian, the contact discontinuity is never smeared out 
at all once it is formed. However, this is not a serious problem in 
the simulation of the 
realistic core collapse, since even the contact surface between the
iron core and the mantle spreads over a considerable distance and
is not a strict discontinuity.  
\par
The time steps for these calculations are 
controlled so that the maximum variation of any dependent variables 
does not exceed 2\%. As a result, the Courant number defined earlier 
is kept  as large as 0.3 through the calculations. The
calculation which allows the maximum variation of 15\% has also been
done in order to see the effect of time step control. The result shows 
the tendency to smear out the shock and the rarefaction wave. However,
the effect is not so large in the non-relativistic case. This issue will be
raised again in the next section. 

\subsection{The Special Relativistic Shock Tube Problem}

Next shown are the numerical solutions for the following Riemann problem:
\begin{equation}
\left \{
\begin{array}{ccl}
\rho_{L} & = & 0.2\\
p_{L} & = & 2 \times 10^{18}\\
U_{L} & = & 0
\end{array}
\right .
\quad \quad
\left \{
\begin{array}{ccl}
\rho_{R} & = & 0.1\\
p_{R} & = & 1 \times 10^{18} \\
U_{R} & = & 0
\end{array}
\right .
\quad \quad ,
\end{equation}
where the units are again CGS. This parameter set is the same as that
used by 
Swesty. We use 100 zones with an equal spatial resolution also in this
case. The equation of state is again $p = (\gamma - 1) \rho_{b} \varepsilon$. 
\par
Figure~{2} shows the distributions of density, velocity, pressure and
specific internal energy density at the time of $3.03 \times 
10^{-10}$sec (490 numerical steps). $\gamma = 5/3$ also in this case.  
As can be seen clearly, the Rankine-Hugoniot condition at the shock
wave is satisfied very well. No oscillation occurs in this case,
either, except for the one mesh just ahead of the contact
discontinuity. The cause of this erroneous jump is the same as for the 
non-relativistic shock tube. Since the initial discontinuity of the density
is smaller in this case than in the previous case, the resulting
error becomes smaller. One thing to be mentioned here is that the
upper limit of the variation of the dependent variables is chosen to be 0.5\%
in this calculation so that the shock can be resolved in a few zones. 
The resultant Courant number is $\sim 0.2$. In order to see the effect of time
steps on the resolution of the numerical solution, we have done 
two other simulations varying the upper limit. In figure~3 we show the 
results for the upper limit of 2\%. The Courant number is $\sim 2$
through the integration. As can be clearly seen, both the shock wave 
and the rarefaction wave are smeared out considerably. This trend is
much more remarkable in figure~4, which shows the results for an upper
limit of 10\% and the Courant number is as large as 8 at the end of
the calculation. It should be also noted that in this calculation the
Courant number is set to be 3 from the beginning. From these results 
it is concluded that
although the present implicit scheme is stable for the Courant number
$>1$, in order to keep the resolution of the solution the maximum 
variation of the dependent variables must be chosen so that the
Courant number should be $\sim 0.2$ at the shock wave. However, note 
that this does not mean that this code cannot not
be applied to the realistic simulation of the  collapse-driven 
supernovae, since what we should overcome is not the Courant limit 
at the shock wave but the limit at the nascent neutron star. This issue 
will be discussed again in the section of adiabatic collapse calculation. 

\subsection{Sedov's Point Explosion in a Uniform Medium}

The dimensional parameters which characterize the problem are the
density of the uniform matter $\rho_{0}$ and the total energy of
explosion $E_{\mbox{tot}}$. The
problem, however, can be formulated in the completely non-dimensional 
form. Hence the solutions with different initial parameters are
equivalent to each other. We choose the following parameters for
simplicity:
\begin{equation}
\left\{
\begin{array}{ccl}
\rho_{0} & = & 1 \\
E_{\mbox{tot}} & = & 1 \\
\gamma & = & 5/3 
\end{array}
\right . \quad \quad ,
\end{equation}
where the units are CGS. The computational region is the sphere with 
the radius of 1, and is
covered by 100 equal baryon mass zones. The equation of state is $p =
(\gamma -1) \rho_{b} \varepsilon$. The calculation is initiated by
depositing the total energy in the central  mesh. For numerical
convenience the uniform ambient matter has the specific internal energy
density of $\sim 10^{-3}$cm$^{2}$/sec$^{2}$. This is justified since 
 the ambient specific internal energy density is negligibly small
compared with that of shocked matter. The results are
shown in figure~5. 
The shock is resolved by $2 \sim 3$ meshes also in this case, and the
analytic solution is reproduced quite well in most of the
computational region. In the central $\sim 10$ meshes, however, the 
numerical solution is not very good. This is because the mesh for this
problem is uniform in baryon mass and the spatial resolution is not 
sufficient for the central region. 

\subsection{The Self-Similar Collapse of a $\gamma = 4/3$ Polytrope}

It is well known that a non-relativistic gas with an adiabatic
index of $\gamma = 4/3$ is neutrally stable against gravitational
collapse. Goldreich \& Weber (1980) found a solution of a self-similarly
collapsing gas with $\gamma = 4/3$. Yahil (1983) generalized this solution to
the collapse for different $\gamma$'s. These self-similar solutions
approximate the homologous collapse of the inner core of the
collapse-driven supernovae. Hence it is indispensable to the numerical 
schemes for the collapse-driven supernovae that they can reproduce 
this self-similar collapse. 
\par
The initial condition is made as follows: assuming the polytropic
equation of state $p = K \rho_{b}^{\gamma}$ with $\gamma = 4/3$, we
first solve the Lane-Emden equation. Here we take $K$ as the value for
the relativistically degenerate gas with electron fraction $Y_{e} =
0.45$. The central density is set to be $1.0 \times
10^{8}$g/cm$^{3}$. Then we take out the inner 1M$_{\sun}$ and reduce
the internal energy uniformly so that the total energy becomes 0. This 
reduction factor is nearly equal to the pressure deficit $d$ defined by
Goldreich-Weber as
\begin{equation}
d \equiv \left ( \frac{M_{\mbox{hc}} (= 1.0 \mbox{M}_{\sun})}
{M_{0} \times 1.0449} \right ) ^{\frac{2}{3}} \quad ,
\end{equation}
where $M_{\mbox{hc}}$ and $M_{0}$ are the homologous core mass which
is set to be 1M$_{\sun}$ in this model and the mass of the 
polytrope without pressure deficit, respectively. 
\par
The computational region (1M$_{\sun}$) is covered with 100 equal
baryon mass zones also in this case. It is noted that the equation of state
for simulation is ideal gas like ($p = (\gamma - 1) \rho_{b}
\varepsilon$) with $\gamma = 4/3$, not adiabatic  
($p = K \rho_{b}^{\gamma}$). This is because we want to see how well the
constancy of entropy is maintained in the course of the numerical
integration. 
\par
The results are shown in figure~6, where the time variation is
presented for density, velocity, pressure and the quantity
proportional to entropy: $p/\rho_{b}^{\gamma}$. 
In this figure, the solid lines are just connecting the data for the
same time step. When we pay  attention to the velocity
distribution, it is clear that most parts of the polytrope collapse 
homologously, that is, the infalling velocity is proportional to the
radius. Although it is not the case for the outer few zones, this is
due to the boundary condition which is set by hand and is not
consistent with the actual dynamics. In the actual simulation of the
supernovae, this problem concerning the outer boundary condition
should be avoided
by locating the boundary at a sufficiently distant position. On the 
other hand, it is also found that the entropy distribution remains 
almost constant for most parts of the polytrope through the
evolution. The exception is the outer few zones also in this case. 
Ignoring these outer parts, we can find the maximum error of the
entropy is about 8\% at the innermost zone and about 5\% for the other 
zones. The calculation is terminated when the central density reaches
$4.85 \times 10^{12}$g/cm$^{3}$ after 600 time steps. The relativistic 
correction is $g_{00} \sim 0.97$ at that time.

\subsection{Oppenheimer-Snyder's Dust Collapse Problem}

The general relativistic collapse of uniform pressureless matter was
analytically solved by Oppenheimer \& Snyder. When we assume the matter 
is initially ($\overline{t} = 0$) at rest with the uniform
density $\rho_{b}{}_{0}$ and the radius $r_{0}$, then the subsequent
motion is described parametrically as follows:
\begin{eqnarray}
\label{eqn:time}
r & = & \frac{r_{0}}{2} (1 + \cos \eta) \\
\overline{t} & = & \frac{1}{2} 
\sqrt{\frac{3}{8 \pi G \rho_{b}{}_{0}}} 
(\eta + \sin \eta) \quad ,
\end{eqnarray}
where $\eta$ is varied from 0 to $\pi$. It should be mentioned that the
above $\overline{t}$ is the proper time for each mass
element and should be distinguished from the coordinate time $t$. 
The boundary condition for the coordinate time $t$ is not altered in 
this simulation, they are related to each other by
\begin{equation}
\frac{dt}{d\overline{t}} = e^{-\phi} \quad .
\end{equation}
The time coordinate in figure~7 is the coordinate time $t$ which is
obtained by numerically integrating equation~(\ref{eqn:time}). However, 
the difference is not significant except for the final few
msec. Note also that in this time coordinate it takes infinite 
time for the mass shell to reach the event horizon. However, it is not 
a serious problem, as stated earlier, since the main purpose of this 
code is to study the process to produce a neutron star.
\par
Following Swesty, we take 2M$_{\sun}$ cloud with a uniform density of 
$\rho_{b}{}_{0} = 10^{8}$g/cm$^{3}$ for the initial model. Since the 
present code cannot handle zero-pressure, we use the polytropic 
equation of state $p = K \rho_{b} ^{\gamma}$ with $\gamma = 4/3$ and
negligibly small $K$. This initial configuration is covered with 160
equal baryon mass meshes. The calculation is
terminated after 3000 time steps when the central density reaches
$1.72 \times 10^{14}$g/cm$^{3}$.
\par
In figure~7, the motions of the representative mass elements are
shown. The solid lines are exact solutions in this case.   
As shown in this figure, the coincidence between the numerical
solution and the exact solution is satisfactory. The slight difference 
(less than 1\% in time) appears only for the last few msec which is
not clear in figure~7. This is partially due to the error of the 
integration of the coordinate time (equation~(\ref{eqn:time})). 
The above result along with the next one clearly shows that the
present code can treat the general relativistic gravity correctly.

\subsection{The Hydrostatic Calculation}

This and the next section are devoted to demonstrate the performance
of this code in the context of the collapse-driven supernovae, that
is, it can calculate the hydrodynamics far beyond the dynamical time 
scale after the proto-neutron star is formed in the central region. This is the
crucial point for the purpose of this paper. 
\par
In this section, we take a hydrostatic configuration for the initial
condition for the dynamical calculation, and see whether this initial
state will be maintained or not. For that purpose, we first solve the
Tolman-Oppenheimer-Volkov equations:
\begin{eqnarray}
\frac{dp}{dr} & = & - (\rho + p) \frac{\stackrel{\sim}{m}(r) 
+ 4 \pi r^{3} p}{r^{2}\left(1 - 
\displaystyle{\frac{2 \stackrel{\sim}{m}(r)}{r}}\right)} \\
\frac{d\phi}{dr} & = & \frac{\stackrel{\sim}{m}(r) + 4 \pi r^{3} p}
{r^{2}\left(1 - \displaystyle{\frac{2 \stackrel{\sim}{m}(r)}{r}}\right)} \\
\rho & = & \rho_{b} (1 + \varepsilon) \\
\stackrel{\sim}{m}(r) & = & \int^{r}_{0} \rho (r') 4 \pi r'^{2} dr' 
\quad ,
\end{eqnarray}
where the notation is the same as in the dynamical equations. The
boundary condition for $\phi$ is consistent with the dynamical
counterpart:
\begin{equation}
e^{\phi_{s}} = \left(1 - \frac{2 \stackrel{\sim}{m}_{s}}{r_{s}}
\right)^{\frac{1}{2}} \quad .
\end{equation}
The subscript $s$ in this equation means the core surface values. 
The equation of state is assumed to be polytropic $p = K
\rho_{b}^{\gamma}$ with $K = 1.97 \times 10^{-3}$ in the CGS units and
$\gamma = 2.5$. The central density is taken to be  $\rho_{b}{}_{c} =
4 \times 10^{14}$g/cm$^{3}$. The initial model thus obtained has the
gravitational mass of $\stackrel{\sim}{M}_{s} = 1.72$M$_{\sun}$ and the
radius of $R_{s} = 7.7$km. This approximate neutron star is nearly the 
same as that used by Schinder et al (1988). 
\par
In the dynamical simulation, we assume the same equation of state and
use 100 uniform baryon mass zones. In figures~8 and 9 the trajectories of all
the mass zones are shown with the time variation of the Courant
number. 
The time step is controlled so that all the dependent variables 
should not change more than 2\% at each time step. Figure~8 
shows the mass element motions for the first 10msec. As can be seen,
nothing occurs except for the very little oscillation during the first
1msec. The Courant number is shown with a curved line across the
trajectories. It is found that the time step is increased
monotonically. This is because the matter velocities decrease
monotonically after the first small transient. Figure~9 shows the
mass trajectories for the same model but for a much longer time scale. 
The final time is 20sec, which should be compared with the dynamical 
time scale of the neutron star $\lesssim 1$msec. It is also noted this 
time corresponds to only 300 numerical time steps. 
The final Courant number exceeds $10^{6}$. These results imply that 
this dynamical calculation reproduced the hydrostatic configuration 
correctly and also guarantee that this code treats the general relativity
appropriately.  

\subsection{The Adiabatic Collapse Calculation}

Finally shown are the results of the model calculations of the
adiabatic core collapse. In these simulations, we assume the simplified 
equation of state which approximately reproduces the realistic
simulations with detailed micro physics and neutrino transfer, and
calculate the collapse of the iron core. The phenomenological
equation of state adopted here is what Takahara \& Sato (1982) used for their
parametric research of the dynamics of core collapse. The total
pressure $p_{tot}$ consists of the cold part $p_{c}$ 
and the thermal part $p_{t}$ as follows:
\begin{eqnarray}
p_{tot}(\rho_{b}, \varepsilon_{t}) & = & 
p_{c}(\rho_{b}) + p_{t}(\rho_{b}, \varepsilon_{t}) \\
p_{c}(\rho_{b}) & = & K \rho_{b}^{\Gamma} \\
p_{t}(\rho_{b}, \varepsilon_{t}) & = & 
(\gamma_{t} - 1) \rho_{b} \varepsilon_{t}
\quad ,
\end{eqnarray}
where $\varepsilon_{t}$ is the specific thermal energy density, and 
the adiabatic index $\Gamma$ is assumed to be density
dependent. Actually $\Gamma = 4/3$ for $\rho_{b} \leq 4 \times
10^{9}$g/cm$^{3}$ and $10^{12} \leq \rho_{b} \leq 2.8 
\times 10^{14}$g/cm$^{3}$, and $\Gamma = 2.5$ for $\rho_{b} \geq 2.8 
\times 10^{14}$g/cm$^{3}$. The value of $\Gamma$ for $4 \times
10^{9} \leq \rho_{b} \leq 10^{12}$g/cm$^{3}$ is the model
parameter. Similarly, the thermal stiffness $\gamma_{t}$ is varied from 
model to model. By adjusting these parameters, we can get both
successful explosion models and stalled shock models. In fact, larger
values  help the shock propagation and tend to strengthen the
explosion. The initial models are Woosley's 15M$_{\sun}$ and
35M$_{\sun}$ precollapse models (\cite{wsl90}). In the former model, 
the central 1.32M$_{\sun}$ iron core is covered by 100 non-uniform mass zones
in which the baryon mass of each cell is increased outward. $\Gamma =
1.30$ and $\gamma_{t} = 1.30$ are assumed. This model explodes in a
prompt way.  In the latter model, on the other hand, 
the central 2.0M$_{\sun}$ is covered with 150
non-uniform meshes and $\Gamma = 1.28$ and $\gamma_{t} = 1.25$ are
assumed. The shock is stalled inside the iron core for this model 
as shown below. 
\par
Figures~10 and 11 show all the mass trajectories 
for the 15 M$_{\sun}$ model 
with the time variation of the Courant number. Again I will show the
results in two different time scales. In figure~10, we can see 
the motions of mass elements until the shock reaches the iron core
surface. As stated above, since the shock propagates the entire iron
core without stagnation and some part of the outer core moves outward
at the break out, this model is a successful prompt explosion model. 
It is also noted that the central part becomes almost hydrostatic. 
The data concerning the location of each mass element and the Courant number
 are saved every 10 time steps. As can be seen, the Courant number
increases up to about 500 at the break out. Figure~11 shows the subsequent 
evolution until the calculation is terminated. The final time is
600msec and corresponds to 5000 numerical  time steps. Since the mantle is not 
taken into consideration, the result after shock breaks out is not of physical
significance. However, the calculation was continued in order to see 
how large the Courant number gets. As can be seen, the final Courant
number exceeds 1000. One thing to be mentioned here is that the
Courant number shown here is determined not at the shock wave but at 
the center of the proto-neutron star, which implies that we could 
overcome the severe Courant limit at the center without instability. 
This seems quite encouraging for the purpose of
this project. It should be mentioned, however, that this is partly 
because there is no significant accretion on the proto-neutron 
star in the successful explosion model. Hence it is of greater
importance to  see the time evolution of the stalled shock model.
\par
In figure~12 we can see the motions of the representative mass
elements for the 35M$_{\sun}$ model until 50msec after the core bounce. 
In this case, the shock does not expel the accreting matter, that is, 
the shock is stalled. The inner core becomes nearly hydrostatic also
in this case. However, since the matter is still falling onto the
proto-neutron star continuously, the Courant number cannot become so
large as the previous model as expected.
The subsequent evolution is shown in figure~13. The final time is
900msec after the start of the simulation. The total number of numerical time steps
is 8000. As can be seen, in this case the Courant number also becomes 
as large as 1000. Again this is a very encouraging result for this 
project. However, it should be mentioned that the resolution of the
outer core is not sufficient to resolve the shock wave for the 
35M$_{\sun}$ model. This is not only because the variation of dependent
variables is allowed up to 2\% in this calculation (see the results of 
the relativistic shock tube problems), but also because matter continues
to concentrate to the center. Hence we must implement some rezoning 
routine or we should use a larger number of zones. This is the problem 
to be solved in the next step. 
  
\section{SUMMARY}

The implicit general relativistic hydrodynamical scheme is coded using
the approximate Riemann solver and extensive numerical tests are
accomplished. The scheme is very simple since the formulae to evaluate
the velocity and the pressure at the cell interface have such simple
forms that we can easily calculate their derivatives with regard to the
dependent variables. The numerical solutions obtained by this scheme
show no oscillation after both shock waves and rarefaction waves
unlike the schemes requiring an artificial viscosity, although 
a small erroneous jump appears after the contact surface. The
calculations of the simple adiabatic collapse of massive stellar 
cores indicate this code is quite stable even if we take a much longer 
time step than the Courant time step, which fact is very encouraging 
for its application to the more realistic calculation of the late time 
explosion. However, it is also found that the time step should be
adjusted so that the Courant number at the shock wave should be $\sim
0.2$. It is also probable some automatic rezoning routine should 
be implemented to resolve the shock wave sufficiently all through the 
calculation.
\par
The purpose of this project is to provide a tool to study the
radiation hydrodynamics of supernova cores far beyond the dynamical
time scale of the neutron star. We are now proceeding to the next step, in 
which the present hydrodynamic code is combined with a neutrino
transfer code using the multi-energy group flux-limited
diffusion scheme. However, the final goal of this project is
to make a Boltzmann solver and combine it with this implicit
hydrodynamical code thereby removing uncertainties regarding the
neutrino transfer. 
\par
In recent years, the importance of multi-dimensional effects on the
supernova explosion has been pointed out and many detailed numerical
simulations have been done. At present, it seems that the clues to the
successful supernova explosion are such multi-dimensional effect and
neutrinos. Although some effort has been made to do the 
simulation of multi-dimensional radiation transfer, it seems still far
from satisfaction. Hence the one-dimensional simulations with detailed
neutrino transfer and multi-dimensional simulations will play
complementary roles to one another. It is also possible that
one-dimensional simulation would serve as a guide to make some approximate 
treatment of the multi-dimensional neutrino transfer. 
 
\acknowledgments

I am grateful to S. Woosley for providing us with the details of his
progenitor models. I also thank H. Suzuki for helpful advice and 
comments and W. Hillebrandt, E. M\"{u}ller and E. Skillman 
for their kind review of the manuscript. The numerical calculations
were mainly done on the
workstations of the theoretical astrophysics group of the University of
Tokyo, the national institute for high energy physics (KEK) and
Max-Planck-Insitut f\"{u}r Astrophysik. This work was partially
supported by the Grant-in-Aid for Scientific Research from
the ministry of Education, Science and Culture of Japan Nos. 05243103,
04234104, 07740344 and by Japan Society for the Promotion of Science (JSPS)
Postdoctoral Fellowships for Research Abroad.

\newpage
\figcaption{The distribution of density, velocity, pressure and specific 
internal energy density at $t = 0.675$sec (400 time steps) are shown
with the analytic solutions (solid lines) for $\gamma = 5/3$.}
\figcaption{The distribution of density, velocity, pressure and specific 
internal energy density for the special relativistic shock tube
problem described in the text at $t = 3.03 \times 10^{-10}$sec 
(490 time steps) are shown
with the analytic solutions (solid lines) for $\gamma = 5/3$.}
\figcaption{The numerical solution for the same Riemann problem as Fig.~2 
but with larger time step (the Courant number $= 2.0$). 
$t = 2.95 \times 10^{-10}$sec (60 steps).}
\figcaption{The numerical solution for the same Riemann problem as Fig.~2 
but with much larger time steps (the Courant number reaches $\sim 8.0$ 
at the end of the calculation). $t = 2.80 \times 10^{-10}$sec (16 steps).}
\figcaption{The numerical solution for Sedov's point explosion
problem described in the text. The time is 0.49sec and the step number 
is 10000. The solid lines are the exact solutions. $\gamma = 5/3$.}
\figcaption{The numerical solution for the self-similar collapse of
a $\gamma = 4/3$ polytrope. The distributions are shown every 100
steps. The times are 0.463, 0.700, 0.804, 0.881, 0.920, 0.937
seconds. The solid lines are just connecting the data for the same time 
step. }
\figcaption{The trajectories of the representative mass elements for the
Oppenheimer-Snyder dust collapse problem are
shown with white circles. The solid lines are exact solutions. The
baryon masses for the trajectories are 0.125, 0.375, 0.625, 0.875,
1.125, 1.375, 1.625, 1.875M$_{\sun}$.}
\figcaption{The mass trajectories for the hydrostatic problem described
in the text. The curved lines across the trajectories show the time
variation of the Courant number. The evolution for only the first 
10msec is shown.}
\figcaption{The same as Fig.~8 but for the whole calculation.}
\figcaption{The mass trajectories for the adiabatic collapse of the 
15M$_{\sun}$ model. The line across the trajectories show the time
variation of the Courant number. The evolution only for the first 
170msec is shown.}
\figcaption{The same as Fig.~10 but for the whole calculation.}
\figcaption{The mass trajectories for the adiabatic collapse of the 
35M$_{\sun}$ model. The line across the trajectories show the time
variation of the Courant number. The evolution only for the first 
400msec is shown.}
\figcaption{The same as Fig.~12 but for the whole calculation.}
\end{document}